# Giant Nernst effect in the crossover between Fermi liquid and strange metal

Yusen Yang[1,*], Qian Tao[2,*], Yuqiang Fang[3,*], Guoxiong Tang[1], Chao Yao[1], Xiaoxian Yan[1], Chenxi Jiang[2], Xiangfan Xu[4], Fuqiang Huang[3,§], Wenxin Ding[5], Yu Wang[6], Zhiqiang Mao[6], Hui Xing[1,†], and Zhu-An Xu[2,#]

[1] Key Laboratory of Artificial Structures and Quantum Control, and Shanghai Center for Complex Physics, School of Physics and Astronomy, Shanghai Jiao Tong University, Shanghai 200240, China

[2] Zhejiang Province Key Laboratory of Quantum Technology and Device, Department of Physics, Zhejiang University, Hangzhou 310027, China

[3] State Key Laboratory of High-Performance Ceramics and Superfine Microstructure, Shanghai Institute of Ceramics, Chinese Academy of Science, Shanghai 200050, China

[4] Center for Phononics and Thermal Energy Science, China–EU Joint Lab for Nanophononics, School of Physics Science and Engineering, Tongji University, Shanghai 200092, China

[5] School of Physics and Optoelectric Engineering, Anhui University, Hefei, Anhui Province, 230601, China

[6] Department of Physics and Materials Research Institute, Pennsylvania State University, University Park, Pennsylvania 16802, USA

* These authors contributed equally to this work;
e-mail: §huangfq@mail.sic.ac.cn, †huixing@sjtu.edu.cn, #zhuan@zju.edu.cn.

The strange-metal state is a crucial problem in condensed matter physics highlighted by its ubiquity in almost all major correlated systems[1-7]. Its understanding could provide important insight into high-$T_c$ superconductivity[2] and quantum criticality[8]. However, with the Fermi liquid theory failing in strange metals, understanding the highly unconventional behaviors has been a long-standing challenge. Fundamental aspects of strange metals remain elusive, including the nature of their charge carriers[1]. Here, we report the observation of a giant Nernst response in the strange-metal state in a two-dimensional superconductor 2M-WS$_2$. A giant Nernst coefficient comparable to the vortex Nernst signal in superconducting **cuprates, and its high sensitivity to carrier mobility, are found when the system enters the strange-metal state from the Fermi liquid state. The temperature and magnetic field dependence of the giant Nernst peak rule out the relevance of both Landau quasiparticles and superconductivity. Instead, the giant Nernst peak at the crossover indicates a dramatic change in carrier entropy when entering the strange-metal state. The presence of such an anomalous Nernst response is further confirmed in other iconic strange metals, suggesting its universality and places stringent experimental constraints on the mechanism of strange metals.**

One major milestone in condensed matter physics is Landau's phenomenological Fermi-liquid (FL) theory dictating the low-temperature behaviors of most metals[9,10]. It maps the low-energy excitations of interacting fermions to the solvable theories of quasiparticles with a set of renormalization parameters carrying the effect of interactions. These quasiparticles are fermions adiabatically "dressed" by particle-hole pairs and serve as charge carriers, constituting the key notion in the modern understanding of metallicity. A prerequisite for a coherent quasiparticle picture is a quasiparticle lifetime $\tau$ long enough so that its decay rate ($\Gamma = \frac{1}{\tau}$) is smaller than that set by its excitation energy $\epsilon$, namely $\hbar\Gamma < \epsilon$. While this requirement on scattering rate is satisfied in FL, it breaks down in systems with increased electron correlation, known as non-Fermi liquids[11]. As the most prominent example in this class, strange metal garners great attention due to its ubiquitous existence in nearly all major correlated systems, covering the normal state of high-$T_c$ cuprates[1,2,12], ruthenates[4,13], iron pnictides[3,14], heavy fermions[5,8], twisted bilayer graphene[6], and even bosonic system[7]. The relevance to high-$T_c$ superconductivity and quantum criticality makes the strange metal one of the most outstanding issues in modern condensed matter physics.

The hallmark of a strange metal is a persisting linear-in-temperature resistivity in stark contrast with the $T^2$ dependence in FL. The corresponding scattering rate is proportional to the temperature and follows a universal relation $\Gamma_{tr} \sim \frac{k_B T}{\hbar}$, which is the so-called Planckian limit[15-17], the highest scattering rate allowed by quantum mechanics. The striking presence of this Planckian dissipation, though not fully understood, is clearly beyond the coherent quasiparticle picture. More interestingly, a surge of recent effort reveals surprising signatures of coexistence of coherent quasiparticles and 'Planckian' dissipators[4,18-20], making the charge carrier problem in strange metals even more puzzling. To unveil intrinsic behavior in strange metals, a promising approach is to track the carriers by transport, spectroscopy or otherwise, across the quasiparticle coherence temperature $T_{FL}$, from the well-understood FL state to the strange-metal state of interest. A transport-entropy sensitive probe is preferred for addressing these critical issues.

Nernst effect in solids is a transverse electric field ($E_y$) generated by a longitudinal thermal gradient ($-\nabla T \parallel x$) under a magnetic field ($B \parallel z$). It depends on two Onsager flows: the transverse entropy flow and the longitudinal charge carrier flow, and measures their relative ratio[21,22]. As a result, it is inherently susceptible to the entropy of charge carriers in solids. Nernst effect has been remarkably successful in probing various intricate energy scales due to mobile quasiparticles, fluctuating Cooper pairs, Abrikosov vortices, and density waves. Most notably, for unconventional superconductors, Nernst effect was proven to be a powerful probe not only for the superconducting transition but also for the vorticity in the pseudogap regime[23,24] and the stripe order[25], providing essential insight into high-$T_c$ superconductivity.

In this report, we explore the carrier entropy change in the strange-metal state of a superconducting 2M stacking WS$_2$ using Nernst measurement. Distinct signatures in the Nernst signal are identified upon the crossover between the FL state and the strange-metal state. Most striking is a giant Nernst peak with a magnitude up to 3.7 µV·K$^{-1}$·T$^{-1}$ right at the quasiparticle coherence temperature $T_{FL} \sim 25$ K, with the peak magnitude strongly dependent on carrier mobility. Such a Nernst response is unexpected in a conventional quasiparticle picture and strongly indicates a significant change of entropy per carrier upon entering the strange-metal state. The Nernst response is further confirmed in other iconic strange metals, including Sr$_2$RuO$_4$, implying its universality, which places stringent experimental constraints on the mechanism of strange metals.

The material we focus on is a two-dimensional van der Wass system, 2M-WS$_2$. It is isomeric to its sibling 2H-WS$_2$, with a different manner of layer stacking[26]. The system is superconducting at a considerably high $T_c$ of 8.8 K among transition metal dichalcogenides and according to a first principles calculation should host topological surface states[27]. The combination of both superconductivity and topological electronic state yields the possibility of topological superconductivity, which has been a subject of extensive effort[28]. Here we focus on the intermediate temperature region of its phase diagram. As shown in Fig. 1(a), while the longitudinal resistivity $\rho$ decreases upon cooling as expected for metals in general, the Hall coefficient $R_H$ (defined as the off-diagonal $\rho_{yx}$ divided by the magnetic field $B$) increases with decreasing temperature and shows a broad peak feature, signaling unusual transport behaviors. Strong anisotropy in transport is also identified in the longitudinal resistivity. Resistivity along the tungsten atom chain direction (∥ $b$ axis), denoted as $\rho_b$, is more than one order of magnitude smaller than that perpendicular to the tungsten chain direction (in-plane and parallel to the $c$ axis, denoted as $\rho_c$). These together yield carrier mobility of one order of magnitude higher for transport along the tungsten chain direction than that perpendicular to the chain direction.

Distinct regimes are identified through the temperature dependence of the carrier scattering rate $\Gamma$, which is extracted from the longitudinal conductivity ($\sigma_{xx} = \frac{ne^2}{m^*} \Gamma_{tr}^{-1}$) and the transverse Hall conductivity ($\tan\theta_H = \frac{\sigma_{xy}}{\sigma_{xx}} = \frac{Be}{m^*}\Gamma_H^{-1}$) independently, where $\theta_H$ is the Hall angle, $\sigma$ the conductivity tensor, $n$ the carrier density, $m^*$ the carrier effective mass, $B$ the magnetic field, and $\Gamma_{tr}$ and $\Gamma_H$ are the transport and Hall scattering rates, respectively. Three regimes are evident in Fig. 1(c, d). The first is on the low-temperature side, where both resistivities exhibit $T^2$ dependence, conforming to standard FL behavior. At approximately between 25 and 120 K, a second regime is marked by a $T$-linear $\Gamma_{tr}$ and a $T$-square $\Gamma_H$. The presence of two distinct $T$ dependence of the two-carrier lifetime in longitudinal and transverse transport is a hallmark of a strange metal[29-33]. The third regime at $T > 200$ K, similar to the strange-metal phase, shows a $T$-linear $\Gamma_{tr}$ and a $T$-square $\Gamma_H$, with a different slope and intercept. A remarkable resemblance between our experimental data and theoretical calculations based on large-$U$ Hubbard and $t$-$J$ models is shown in Fig. 1(b) [34-37]. The calculation revealed several regimes separated by crossovers in between. The first two regimes are FL and strange metal, as we already identified in the experiment from the well-defined temperature dependence of the scatterings. The third is an empirical high-temperature bad-metal state, which similar to the strange-metal state, does not host well-defined quasiparticles but carries qualitative differences, such as the different slope in the temperature-dependent resistivity. The crossover from FL to strange metal at the quasiparticle coherence temperature $T_{FL}$ is marked explicitly in Fig. 1(c, d).

The entropy transport in these three phases probed by the diagonal and off-diagonal thermoelectric responses are presented in Fig. 2. The Seebeck coefficient, which is the diagonal component of the thermoelectric tensor $S_{xx}$, shows a complex behavior. The Seebeck coefficient along the $b$ aixs, denoted as $S_b$, exhibits weak temperature dependence at high temperatures, a steep drop upon cooling at around 70 K, a subsequent sign change, and a peak at lower temperatures. $S_b$ reduces towards zero at low temperatures, as expected for any entropic process.

In the FL regime, Seebeck coefficient $S = -\frac{\pi^2}{3}\frac{k_B^2 T}{e}\left[\frac{\partial \ln \tau(\epsilon)}{\partial \epsilon} + \frac{\partial \ln N(\epsilon)}{\partial \epsilon}\right]_{\epsilon=\epsilon_F}$ reflects both the band curvature and the energy dependence of scatterings[38], where $k_B$, e, $N$, $\epsilon$, $\epsilon_F$ denote Boltzmann constant, electron charge, electronic density of states, electronic energy, and the Fermi energy, respectively. In light of the overall hole-type carrier in the entire temperature range probed by the Hall coefficient, the sign change in Seebeck suggests significant energy dependence of charge transport. On the other hand, the Seebeck coefficient measured along the c aixs, denoted as $S_c$, shows a quantitatively different temperature profile, demonstrating significant anisotropy in transport. Here, we emphasize that the peak in $S_b$ exists right at $T_{FL}$, an intriguing signature associated with the unconventional entropy transport in the crossover between FL and strange metal.

The off-diagonal thermoelectric response, known as the Nernst signal $N = S_{yx}$, exhibits instead a much clearer behavior. As shown in Fig. 2(c, d), the Nernst signal along the b axis, denoted as $N_b$, exhibits a single peak profile with the peak temperature right at $T_{FL}$. Most notably, the magnitude of $N_b$ is giant. The corresponding Nernst coefficient $\nu_b$, defined as $\nu_b = N_b/B$, is up to 3.7 µV·K$^{-1}$·T$^{-1}$, which is comparable to the highest vortex Nernst response in cuprate superconductors, one of the largest Nernst sources[21,24]. Along with this observation, the Nernst signal $N_b$ is more than twenty times larger than the corresponding Seebeck coefficient $S_b$, meaning that a longitudinal thermal gradient generated an almost entirely transverse thermoelectric voltage, which implies an unconventional underlying mechanism. In stark contrast, the thermoelectric transport measured with $-\nabla T$ along the orthogonal in-plane direction is rather different. $N_c$, the Nernst response for $-\nabla T$ along the c axis, shows a similar single-peak profile, with a lower peak temperature of ~ 14 K. The magnitude of $N_c$ is much smaller, with the coefficient $\nu_c$ only about 0.05 μV·K⁻¹·T⁻¹. The relative magnitude between $N_c$ and $S_c$ reduces dramatically to less than 1, a typical value seen in many conventional cases.

The relationship between Seebeck and Nernst responses is well understood in isotropic one-band systems. As depicted schematically in Fig. 2(a, c), the Nernst signal can be decomposed as $N = -S \tan\theta_H + \frac{\alpha_{xy}}{\sigma_{xx}}$, where $\alpha_{xy}$ is the off-diagonal Peltier tensor, which is proportional to the energy dependence of the Hall conductivity, namely $\alpha_{xy} \sim \frac{\partial \sigma_{xy}}{\partial \epsilon}$. It means that the Nernst response is the residual of two canceling currents, one conventional Hall current $-S \tan\theta_H$, and the other the Peltier Hall current $\frac{\alpha_{xy}}{\sigma_{xx}}$. When the scattering is energy-independent, this leads to a vanishing Nernst response, known as the Sondheimer cancellation[39,40]. The Hall current term $-S \tan\theta_H$ can be evaluated by independent Hall and Seebeck measurements and were shown by the red lines in Fig. 2(c, d). Clearly, for both $N_b$ and $N_c$, the conventional Hall current term accounts for only a tiny fraction of the Nernst signal, which implies an unusual source related to the energy-dependent Peltier term.

With finite energy dependence in the carrier scatterings, the Nernst response in a FL state is set by the ratio between the quasiparticle mobility and the Fermi energy $E_F$ [38], namely

$$\nu = \frac{\pi^2}{3}\frac{k_B^2 T}{e}\frac{1}{B}\frac{\partial \tan\theta_H}{\partial \epsilon}\bigg|_{\epsilon=\epsilon_F} \sim \frac{\pi^2}{3}\frac{k_B^2 T}{e}\frac{1}{B}\frac{\tan\theta_H}{\epsilon_F} = \frac{\pi^2}{3}\left(\frac{k_B}{e}\right)\frac{k_B T}{\epsilon_F}\mu_H \qquad (1)$$

In eq. 1, the product of the temperature $T$ and the mobility $\mu_H$ warrants a peak profile in Nernst at low temperatures. The blue curves in Fig. 3(a, b) represent this FL contribution estimated from eq. 1 using extracted mobility[26] and $E_F$ values reported earlier[27]. Apparently, for $-\nabla T \parallel c$, eq. 1 provides an adequate description of the experimental data, with consistent magnitude, overall temperature dependence, and the peak temperature at 14 K. This suggests that the single-band picture describes remarkably well the behavior in 2M-WS₂. In contrast, for $-\nabla T \parallel b$ the FL contribution $\nu_{b,\text{FL}}$ constitutes only a small fraction of the experimental Nernst coefficient and exhibits a peak at a much lower temperature of ~12 K. An anomaly in the temperature derivative of the Nernst coefficient $\frac{d\nu_b}{dT}$ appearing right at the same characteristic temperature justifies the FL component in the total Nernst coefficient. Clearly, there is an extra term in the Nernst coefficient with an origin beyond the FL picture, hence the total Nernst coefficient decomposes as $\nu = \nu_{\text{FL}} + \nu'$. Indeed, for both $\nu_b$ and $\nu_c$, we identify consistently $\nu_{FL}$ with its peak temperature set by the mobility and the Fermi energy and $\nu'$ with a peak temperature at $T_{\text{FL}}$ [26]. $\nu'$ is anomalously large in $\nu_b$ but reduces by two orders of magnitude in $\nu_c$, hinting a strong correlation with the carrier mobility. The strong dependence of the amplitude of $\nu'$ on mobility points to a mechanism that is fragile against external scattering. The key question is what the $\nu'$ term corresponds to. In the inset of Fig. 3(a), the profile of $\nu'$ for $\nabla T \parallel b$, after subtracting the $\nu_{FL}$ from the total Nernst coefficient $\nu$, maintains a similar $T$-dependence with that of $\nu$ since the Fermi-liquid term is very small.

No documented source of large Nernst effect explains the present case. The Nernst signal corresponding to the mobile Abrikosov vortex or the superconducting fluctuation is excluded as the present Nernst coefficient is nearly field-independent and persists up to a field of 14 T, which is high enough to suppress the superconducting order in this system. The multi-band physics is another plausible mechanism that often gives rise to enhanced Nernst signal. However, our analysis in Fig. 3 shows that a single-band FL picture has provided an adequate account for the low-temperature thermoelectric behavior with consistent characteristic temperatures identified in the experimental data, which discounted the possibility of the multi-band physics. The nearly one order of magnitude larger Nernst coefficient compared to that estimated using an FL picture implies an unconventional origin. We emphasize that the giant peak in $\nu'$ at the temperature $T_{\text{FL}}$ signifies a direct correlation with the FL to strange metal crossover, which is demonstrated in the two contour plots in Fig. 4(a, b) over a wide range of magnetic fields. Upon increasing $T$ across the crossover temperature $T_{FL}$, the charge carriers transform from well-defined quasiparticles in FL to fermionic excitation in strange metal whose nature is so far not clear. In view of the entropy transport, the quasiparticles in FL with a long lifetime (characterized by $\frac{\hbar}{\tau} \ll k_B T$) carry entropy that is determined by the band structure near the Fermi surface. It is precisely the entropy per carrier that is probed by the Nernst signal, namely $N = \frac{J^S}{J^e}$, according to the Onsager's reciprocity[21]. This argument is purely based on thermodynamics and does not assume a particular type of charge carriers. In our experiment, the anomalously large Nernst coefficient right at the crossover from the quasiparticles in FL to the elusive fermionic excitation in strange metal indicates a significant change in the charge carriers' entropy. Moreover, the identification of both a coherent quasiparticle response and an anomalous response of carrier in the strange-metal phase in our Nernst measurement provide support for the possible coexistence of two types of carrier excitations[19,20].

It is instructive to examine the Nernst response in known strange metals with FL ground state to see whether the anomalous response in 2M-WS$_2$ is a lone anomaly. The cuprate and ruthenate superconductors are the two most iconic cases. The absence of any trace in Nernst response in high-$T_c$ cuprates at $T_{FL}$ [24] is expected as the mobility in high-$T_c$ cuprates is relatively small (typically 1-10 cm$^2\cdot$V$^{-1}\cdot$s$^{-1}$) such that $\nu'$ is too small to be resolved. In Sr$_2$RuO$_4$, the mobility is significantly higher (>100 cm$^2\cdot$V$^{-1}\cdot$s$^{-1}$), making the observation of $\nu'$ plausible. Indeed, we found a well-defined anomaly in its Nernst response right at $T_{FL} \sim 20$ K in this system[41]. As shown in Fig. 4(c), the remarkable resemblance between the two different and independent systems further corroborates our findings. A possible correlation between the Nernst coefficient at $T_{FL}$ and the carrier mobility is summarized in Fig. 4(d). Instead of the linear dependence on the mobility in Fermi liquids, the Nernst coefficient in the strange-metal state exhibits a much stronger dependence on the carrier mobility which again testifies the unusual carriers in strange metals[26]. While a comprehensive theoretical framework for describing the carriers in strange metals is not available, quasiparticle fractionalization[42] or particle-hole asymmetric excitations[43] in strange metals are among the most attractive possibilities. Our observation of significant entropy change of carriers imposes stringent experimental constraints on the physics of strange metal.

**Methods:**

The details of the growth of single-crystalline 2M-$WS_2$ and $Sr_2RuO_4$ samples are provided in Ref. 27 and 44, respectively. Resistivity, Hall, and thermoelectric measurements were performed in a Quantum Design physical property measurement system with a 14-Tesla magnet. A steady-state technique was used in thermoelectric measurements. A temperature gradient, around 0.4 K/mm, was applied in the basal plane and determined by a pair of calibrated differential type $E$ thermocouples.

**Acknowledgments:** We acknowledge useful discussion with Tony Leggett, B. Sriram Shastry, Xiao Lin, Ying Liu, and Hao Zeng. This work was supported by the National Key Projects for Research & Development of China (Grant No. 2019YFA0308602), National Natural Science Foundation of China (Grant No. 11804220, 12174334, 52103353), Natural Science Foundation of Shanghai (Grant No. 20ZR1428900) and the Key Research & Development Program of Zhejiang Province, China (Grant No. 2021C01002).

**Author contributions:** H.X. conceived the project; H.X. and Z.X. supervised the experiments; Y.Y. and H.X. performed the electron transport and thermoelectric transport measurements with assistance from G.T., C.Y., X.Y., Q.T., X.X., and Z.X.; Y.F. and F.H. grew and characterized the 2M-$WS_2$ single crystals; Y.W. and Z.M. grew and characterized the $Sr_2RuO_4$ single crystals; W.D. provided theoretical support; H.X. wrote the manuscript with input from all of the co-authors.

**Competing interests:** The authors declare no competing interests.

**Additional information**

**Supplementary Information** is available for this paper.

**Data availability**

The data that support the plots within this paper and other findings of this study are available from the corresponding author upon reasonable request.

**Correspondence and requests for materials** should be addressed to Hui Xing.

Figure 1

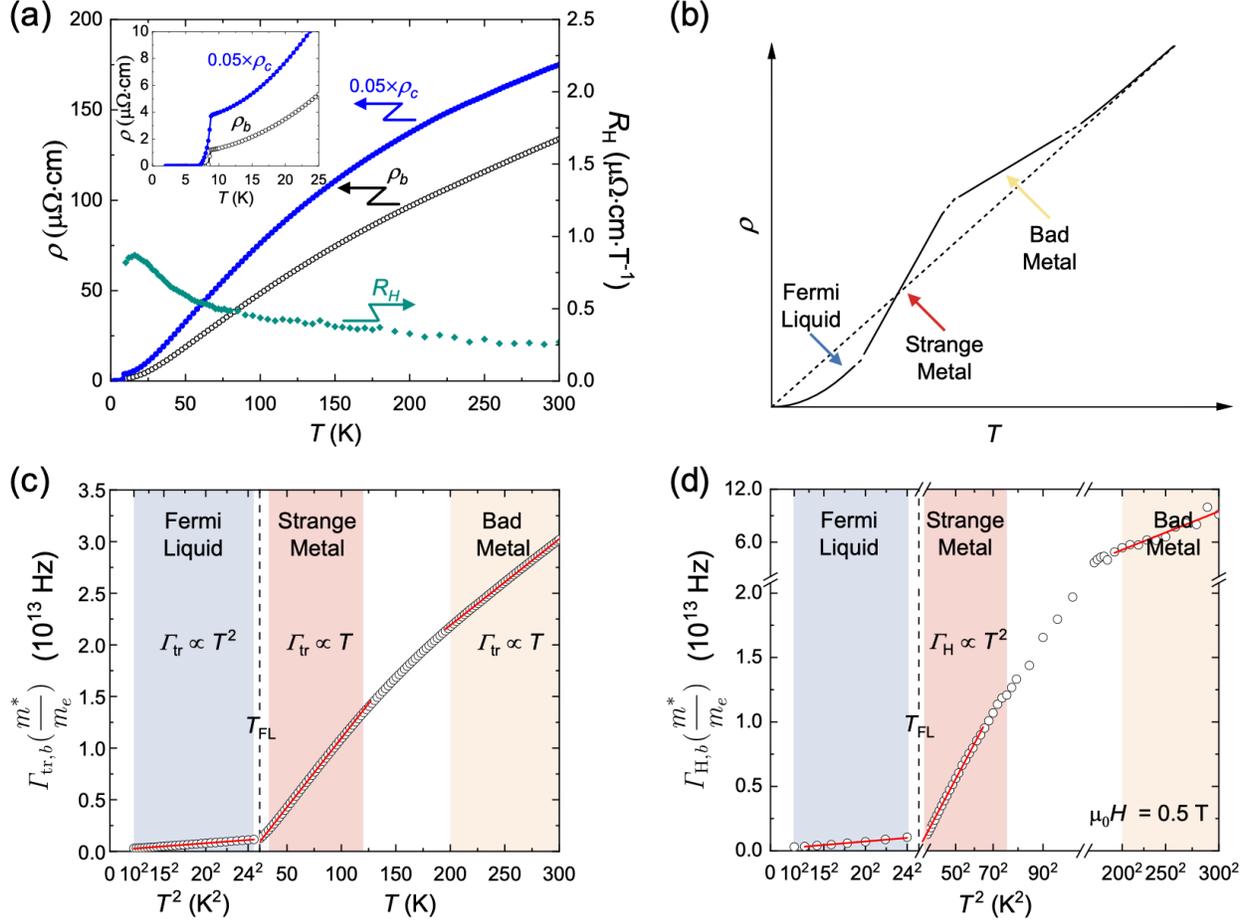

Fig. 1: (a) The temperature dependence of the resistivity of 2M-WS$_2$ for current applied along the in-plane *b* direction (parallel to the W chain direction, denoted as $\rho_b$) and in-plane *c* direction (perpendicular to the W chain direction, denoted as $\rho_c$), respectively. Note that 0.05× $\rho_c$ is plotted to enable better comparison since $\rho_c$ is more than 30 times larger than $\rho_b$. Plotted using the right *y*-axis is the temperature dependence of the Hall coefficient $R_H$ of 2M-WS$_2$. Inset shows the superconducting transition at $T_c \sim$ 8.8 K. (b) A schematic of the different regimes of temperature-dependent resistivity of large-*U* Hubbard and *t-J* models[34-37]. At the lowest temperature, the resistivity is proportional to $T^2$, typical for a Fermi liquid. This behavior terminates at the quasiparticle coherent temperature $T_{FL}$. Upon warming, the resistivity becomes linear-in-temperature and the system enters the strange-metal regime. At higher temperatures, the system is in an empirical bad-metal regime with a resistivity that increases linearly but with a different slope. (c, d) The temperature dependence of scattering time extracted from resistivity and Hall resistivity, denoted as $\Gamma_{\text{tr},b}$ and $\Gamma_{\text{H},b}$, respectively. A factor of $\frac{m^*}{m_e}$ absorbs the correction of effective mass $m^*$, which is not determined in our experiments, where $m_e$ is the bare electron mass. Red lines are linear fittings in different temperature ranges.

Figure 2

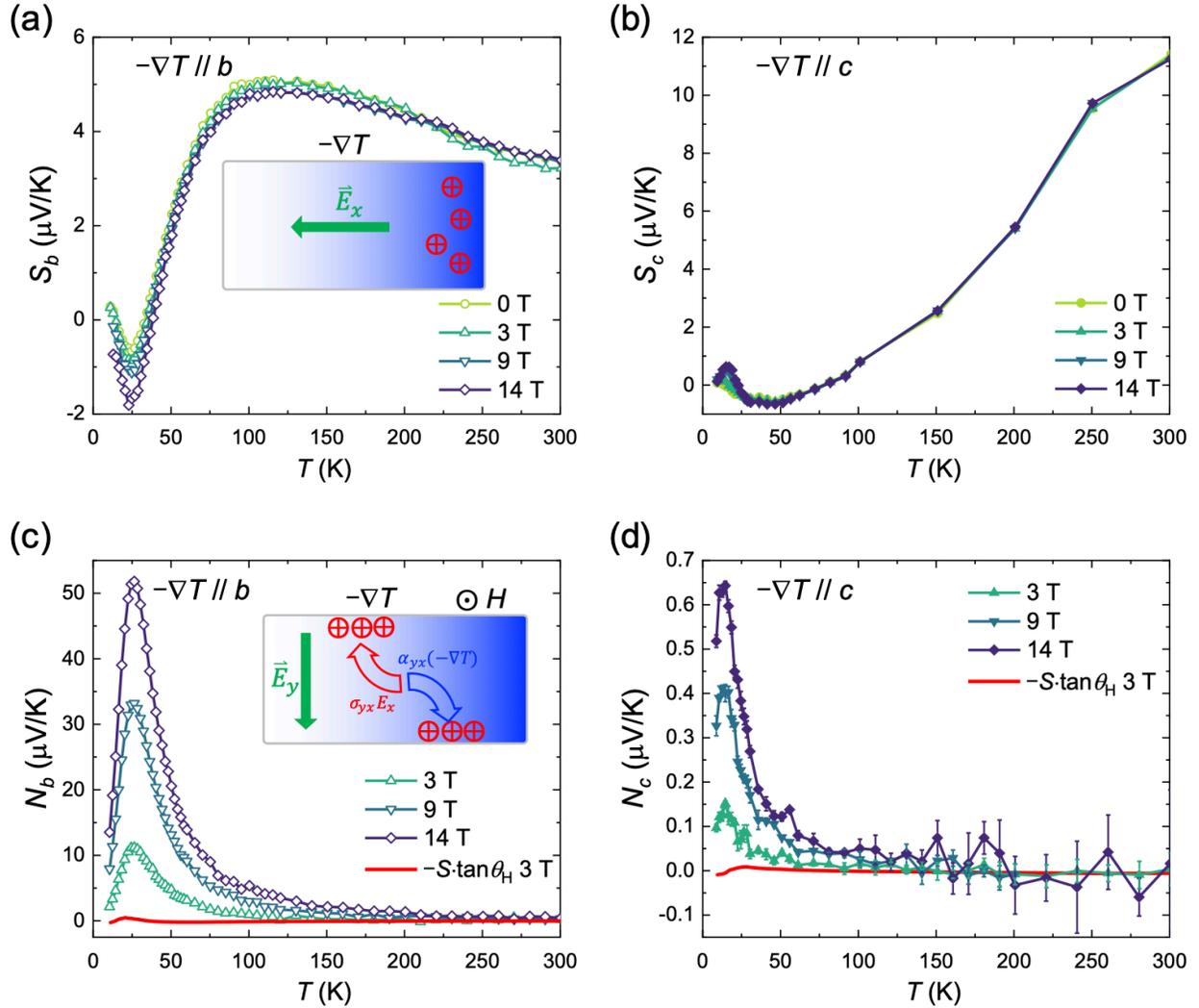

Fig. 2: The temperature dependence of Seebeck coefficient $S$ and Nernst signal $N$ in 2M-WS$_2$ for thermal gradient along the $b$ axis (a, c) and the $c$ axis (b, d), respectively. Insets are schematics showing a simplified picture of the composition of Seebeck and Nernst signals. Red and blue arrows indicate the conventional Hall current $\sigma_{yx}E_x$ and the Peltier Hall current $\alpha_{yx}(-\nabla T)$ and their respective deflection in a magnetic field. Red lines in (c, d) represent the conventional Hall component $-S\tan\theta_H$ at a selected magnetic field of 3 T.

Figure 3

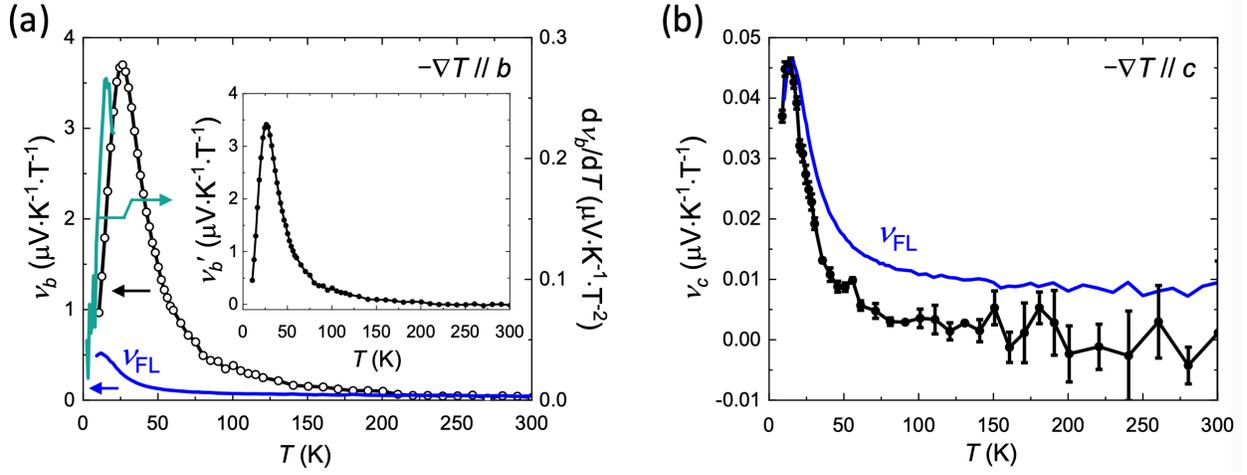

Fig. 3: The temperature dependence of Nernst coefficient $v$ in 2M-WS$_2$ for thermal gradient along the $b$ axis (a) and the $c$ axis (b), respectively. Both are measured in a magnetic field of 14 T. Blue lines are the Fermi-liquid contribution $v_{FL}$ calculated using equation (1). The green line representing the derivative $\frac{dv(T)}{dT}$ shows a peak at a characteristic temperature very close to the peak temperature of Fermi-liquid Nernst component. Inset in (a) shows $v'$ which is the measured Nernst coefficient $v$ subtracted by the Fermi-liquid component $v_{FL}$.

Figure 4

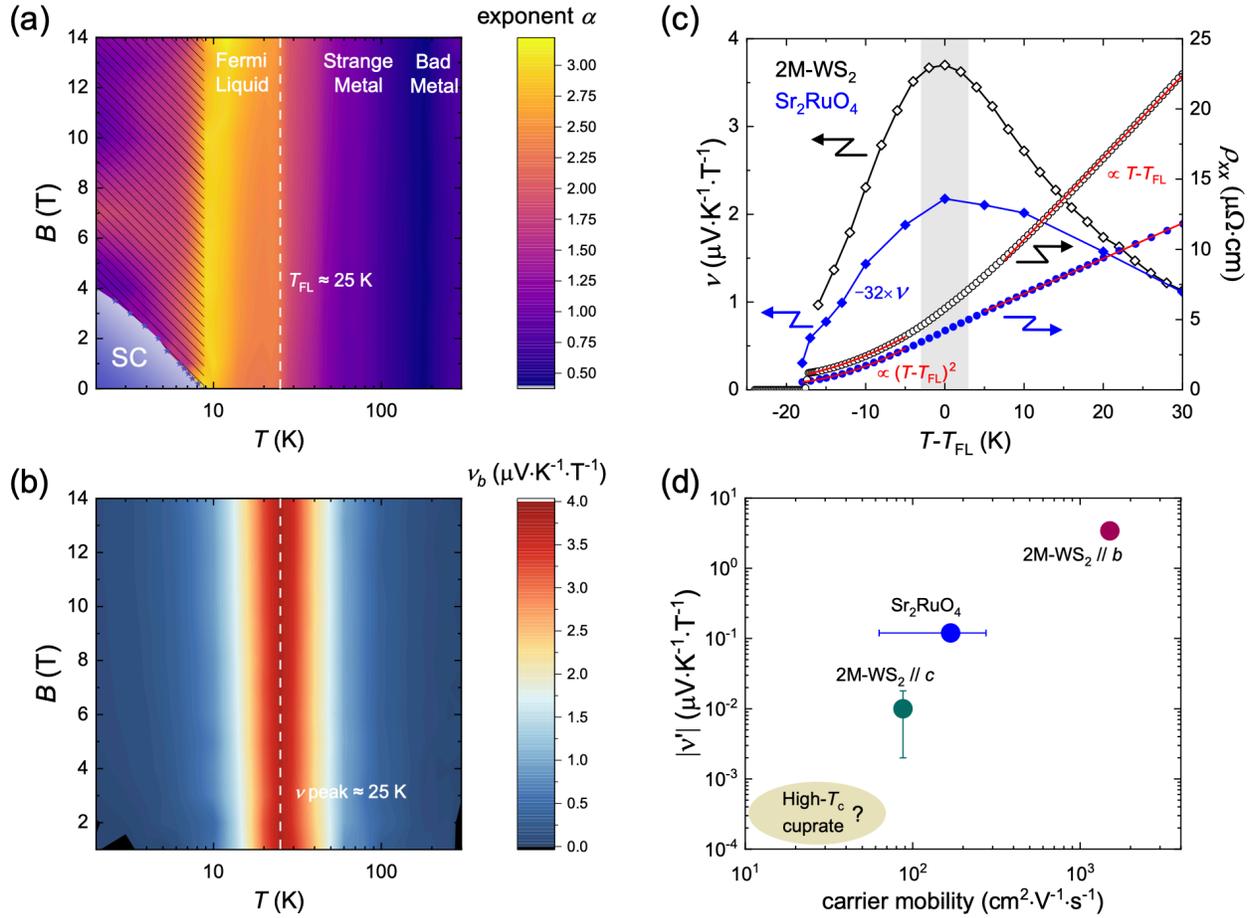

Fig. 4: (a) A contour plot of the temperature and magnetic field dependence of the exponent $\alpha$ in the $b$-axis scattering rate $\Gamma_{\text{tr},b} \sim A' \cdot T^{\alpha}$ in 2M-WS$_2$. (b) A contour plot of the temperature and magnetic field dependence of the Nernst coefficient $\nu_b$ for thermal gradient along the $b$ axis in 2M-WS$_2$. (c) A comparative plot of the temperature dependence of the resistivity and Nernst coefficient in 2M-WS$_2$ and Sr$_2$RuO$_4$. For the Nernst coefficient in Sr$_2$RuO$_4$, $-32\times \nu$ is plotted to enable better comparison. The negative sign compensates for the negative Nernst coefficient in Sr$_2$RuO$_4$ due to its positive Seebeck coefficient in the temperature range of interest, which is opposite to that of 2M-WS$_2$ [26]. Red lines indicate the $T$-linear and $T$-square fittings. (d) The peak magnitude of the anomalous term $\nu'$ in Nernst coefficient in different systems with Fermi liquid to strange metal crossover plotted against their respective carrier mobility. The three dots represent $|\nu'|$ in 2M-WS$_2$ with an in-plane thermal gradient along the $b$ axis, Sr$_2$RuO$_4$ with an in-plane thermal gradient and 2M-WS$_2$ with an in-plane thermal gradient parallel to the $c$ axis. The possible location of high-temperature cuprate superconductors is also indicated.